\documentclass{PoS}

\title{Angular Correlations in STAR}

\ShortTitle{Angular Correlations in STAR}

\author{\speaker{Michael Daugherity for the STAR Collaboration}\\
        University of Texas, Austin, Texas 78712, USA\\
        E-mail: \email{daugherity@physics.utexas.edu}}

\abstract{
The related studies of two-particle correlations and event-by-event fluctuations have 
played important roles in the search for new physics through the experimental study of 
relativistic heavy ion collisions. We present a general method of determining minimum-bias 
two-particle correlations and show the relationship between these correlations and event-by-event 
fluctuations. Data from the STAR experiment at RHIC for Au-Au collisions at $\sqrt{s_{NN}}$ =
 62 and 200 GeV will be presented that show the energy and centrality dependences of angular correlations. 
This analysis provides a unique method for studying the interaction of semi-hard scattered partons 
with the dense medium produced at RHIC as well as properties of the bulk medium itself, 
and will shed new light on the sources of non-statistical fluctuations.
}

\FullConference{The 2nd edition of the International Workshop --- Correlations and Fluctuations in Relativistic Nuclear Collisions --- \\
  July 7-9 2006\\
  Galileo Galilei Institute, Florence, Italy}

\begin{document}

  \section{Introduction}
  
  The analysis of correlations and fluctuations provides essential information on the nature of the medium produced in ultra-relativistic heavy-ion
  collisions.  For example, correlation studies often focus on specific effects such as the collective behavior leading to elliptic flow, 
  quantum correlations, and high-$p_{t}$ 
  jet correlations.  Event-by-event fluctuation analyses of global parameters (\emph{e.g.} mean $p_{t}$, net charge, etc.) 
  search for critical fluctuations as a signature of a phase transition.  
  This dichotomy is evidence of a false perception that correlations and fluctuations are unrelated.
  This work presents a general method of measuring \emph{minimum-bias two-particle correlations},
  where a two-particle correlation is constructed using all detected particles from an event combined into all possible unique pairs, 
  that includes the above mentioned individualized correlation analyses and is directly related to event-by-event fluctuations.

  In Section 2 a correlation measure is derived from standard statistical definitions.  In Section 3 this method is used to measure angular (pseudorapidity $\eta$
  and azimuth $\phi$) correlations in 62 and 200 GeV Au-Au collision data from the STAR detector at RHIC \cite{STAR NIM}.  
  These results, combined with similar analyses
  at 130 GeV and of 200 GeV proton-proton collisions, present a survey of the system energy and centrality dependence of two-particle correlations at RHIC. 
  The evolution with centrality and comparison to proton-proton events and HIJING simulations  indicate 
  that parton scattering and fragmentation are modified in central heavy-ion collisions, presumably through interactions with a dense medium.
  In Section 4 the technique is extended to differentiate between electric charge sign to form charge-dependent correlations which provide differential information 
  on hadronization in contrast to that inferred from net-charge fluctuations. 

  \section{Correlation measures}

  Two-particle correlation measures are formed by comparing the two-particle density $\rho(\vec{p_{1}},\vec{p_{2}})$ to an appropriate reference
  scaled by a normalization factor.  The object of interest is the two-particle density of pairs of particles chosen from the same event, referred to as
  \emph{sibling} pairs, defined as $\rho_{sib}(\vec{p_{1}},\vec{p_{2}})$.  Pairs of particles from different but similar events provide a mixed-event 
  reference $\rho_{ref}(\vec{p_{1}},\vec{p_{2}})$ containing important detector acceptance and phase space effects.  These two factors may be combined in a
  number of ways.  For analysis of quantum correlations (HBT) $\rho_{sib} / \rho_{ref}$ is projected onto a small region of relative momentum space where approximately
  every pair is correlated.  Another common choice is $\frac{\rho_{sib}-\rho_{ref}}{\rho_{ref}}$, where the ratio 
  removes detector acceptance and provides cancellation of systematic effects.  These ratios are \emph{per pair} measures and thus contain a $1/\bar{N}$ 
  dilution factor for average multiplicity $\bar{N}$.  Measures containing such dilution factors are not directly comparable across system energies, centralities,
  incident particle species, or any other condition affecting multiplicity, but more importantly per pair measures can be misleading.
  For example, elliptic flow measure $v_{2}$ includes
  a $\bar{N}^{-1/2}$ scaling, giving the impression that elliptic flow is quite large in peripheral collisions, and therefore does not smoothly extrapolate to the
  proton-proton limit where the $\cos [2(\phi_{1} - \phi_{2})]$ correlation component is negligible.
  
  Two-particle correlations are often measured in the context of a high-$p_{t}$ triggered and associated particle, leading to a model-dependent jet hypothesis
  and trigger bias, however this condition is not necessary.  The same analysis can be performed constructing every possible pair for each event 
  (\emph{i.e.} ``triggering'' on every particle), creating a \emph{minimum-bias}
  correlation having several advantages over a conditional correlation.  
  As particles interact with the medium in RHIC collisions they lose energy, placing particles which
  have significant interactions potentially below the triggered momentum region.  Removing this restriction allows the search for particles with enough
  momentum to be easily detected, yet soft enough to have significant interactions with the medium, conceptually similar to finding an ideal Brownian probe 
  for heavy-ion collisions \cite{pt corr}.
  It is also possible to use correlations to explore the bulk medium itself, such as the response of the medium to a hard scattered 
  parton \cite{pt scaling}.  Finally, most correlation analyses must extract the correlation signal of interest, such as jets, HBT, or flow,
  from other processes, but a minimum-bias analysis gives a complete picture of the dynamics of the system and far more information to use in determining the
  relative contributions from these components.  

  It is possible to derive a correlation measure from standard statistical definitions that retains the useful properties of removing detector acceptance effects
  and the cancellation of tracking inefficiencies and other systematic errors in the ratio $\rho_{sib} / \rho_{ref}$,
  but also scales with N in way that allows for direct comparison between different collision systems, energies, and centralities.  One way to achieve this scaling is
  to create a correlation measure that would be independent of centrality if heavy-ion collisions were linear superpositions of proton-proton collisions, making deviations 
  from this participant number scaling directly accessible.  The canonical definition of a correlation between two
  variables, usually referred to as Pearson's normalized covariance or correlation coefficient, is the ratio of the covariance between the two variables with 
  the product of their standard deviations: $corr(a,b) = \frac{cov(a,b)}{\sigma_{a}\sigma_{b}}$.   
  This work will focus on correlation of the number of pairs of particles, but may 
  easily be extended to other quantities. Event-by-event fluctuation measures can be defined in terms of an integral of $corr(a,b)$, but examining the integrand 
  directly reveals new differential information of the system.
  Only correlations will be shown here,
  but the relationship between fluctuations and correlations is emphasized to place these results in a larger context and highlight the applicability of 
  this analysis method.  A related study \cite{pt scaling} illustrates this by explicitly inverting a transverse momentum fluctuation integral equation to find the 
  associated correlations.  

  Returning to the definition $corr(a,b) = \frac{cov(a,b)}{\sigma_{a}\sigma_{b}}$, we must express densities $\rho_{sib}$ and $\rho_{ref}$ as a covariance.
  Particles from different events must be uncorrelated, so the two-particle density $\rho_{ref}$ can be factorized as a product of one-particle densities.  
  Defining $\Delta\rho = \rho_{sib} - \rho_{ref}$ we see that in terms of particle counts $n$ in histogram bin (a,b) 
  $(\epsilon_{a}\epsilon_{b})\Delta\rho = \overline{n_{a}n_{b}} - \bar{n}_{a}\bar{n}_{b}  = \overline{(n-\bar{n})_{a}(n-\bar{n})_{b}}$, where $\epsilon$ is the bin
  width used to convert densities to bin counts and the overline denotes an average over events.  
  Thus $\Delta\rho$ is a covariance, which is more conveniently expressed as particle counts than densities.  
  In the Poisson limit $\sigma_{a}\sigma_{b} = \sqrt{\bar{n}_{a}\bar{n}_{b}}$, defining our correlation measure as the density ratio: 
  $$ \frac{\Delta\rho}{\sqrt{\rho_{ref}}} \equiv \frac{1}{\epsilon_{a}\epsilon_{b}}\frac{\overline{(n-\bar{n})_{a}(n-\bar{n})_{b}}}{\sqrt{\bar{n}_{a}\bar{n}_{b}}}$$ 
  
  The correlation measures $\Delta\rho$, $\Delta\rho/\rho_{ref}$, and $\Delta\rho/\sqrt{\rho_{ref}}$ have very similar structures in angular 
  (pseudorapidity $\eta$ and azimuth $\phi$) space when corrected for acceptance so that $\rho_{ref}$ is approximately uniform, though with different overall normalizations. 
  Of these, only $\Delta\rho/\sqrt{\rho_{ref}}$ 
  is a \emph{per-particle} measure which allows for direct comparison of correlations from events regardless of multiplicity.  
  The square root in the denominator provides a bin-by-bin normalization rather than applying a global normalization of some factor of $\bar{N}$ across all bins.  
  This distinction is vital in transverse momentum space where both  $\rho_{sib}$ and $\rho_{ref}$ are rapidly varying.

  \section{Experiment}
  Data for this analysis were obtained from the STAR detector at RHIC measuring charged particle production from Au-Au collisions at $\sqrt{s_{NN}}$ of 62 and 200 GeV.
  A minimum-bias event sample required coincidence of two Zero-Degree Calorimeters, a minimum threshold in the Central Trigger Barrel, and a reconstructed
  primary vertex.  
  Particle tracks were reconstructed with the Time Projection Chamber operating within a uniform 0.5 T axial magnetic field. 
  Tracks satisfying standard quality requirements were accepted within full azimuth, $|\eta|\leq1.0$, and $0.15 \leq p_{t} \leq 6.0$ GeV/c.
  Electrons and positrons were suppressed with a dE/dx cut within $0.2 \leq p \leq 0.4$ GeV/c, but no other particle identification was attempted.  
  Pair-wise tracking inefficiencies were corrected by applying track merging and splitting cuts to both sibling and mixed-track pairs.
  To reduce systematic error, events were grouped into multiplicity and primary z-vertex classes such that within a class 
  $\Delta N < 50$ and $\Delta z < 5 $ cm.  Reference pairs were created by only mixing events within
  the same class.  Furthermore, the correlation measure was constructed for each electric charge pair type (\emph{i.e.} $++$, $+-$, $-+$, and $--$) for each
  class, and then combined within each centrality bin.  Events for both energies were divided into 11 centrality bins defined by total cross
  section fraction as 90-100\% (peripheral), 80-90\%, 70-80\%, \ldots, 10-20\%, 5-10\%, and 0-5\% (central).   
  
  The two-particle correlation $\frac{\Delta\rho}{\sqrt{\rho_{ref}}}$ was measured as a function of angular difference variables 
  $\eta_{\Delta} \equiv \eta_{1} - \eta_{2}$ and $\phi_{\Delta} \equiv \phi_{1} - \phi_{2}$.  Correlation structures are found to be invariant along the sum 
  variables $\eta_{1} + \eta_{2}$ and $\phi_{1} + \phi_{2}$, thus projections onto coordinates ($\eta_{\Delta}$, $\phi_{\Delta}$) compactly represent the entire 
  4D angular space ($\eta_{1},\eta_{2},\phi_{1},\phi_{2}$) without loss of information.      

  

  \section{Charge-independent correlations}
  
  To understand two-particle correlations in the complex system of heavy ion collisions, we first examine proton-proton (p-p) collisions as a reference.
  p-p angular correlation data are well described by a two-component model of string plus jet fragmentation, a similar model is often used 
  by event generators such as PYTHIA \cite{pythia}, HIJING \cite{hijing}, and AMPT \cite{ampt} to simulate particle production mechanisms.
  Analysis of the multiplicity 
  dependence of transverse momentum spectra in p-p collisions showed that the spectra can be separated into soft and
  semi-hard components \cite{pp spectra}.  Based on this result, a follow-up study of two-particle correlations observed that the soft and 
  semi-hard components were distinctly separated in transverse rapidity $y_{t}$\footnote{Transverse rapidity 
    $y_{t} \equiv \frac{1}{2}\ln(\frac{E+p_{t}}{E-p_{t}}) = \ln(\frac{m_{t} + p_{t}}{m_{0}})$ 
    with transverse momentum $p_{t}$, transverse mass $m_{t} = \sqrt{p_{t}^2 + m_{0}^2}$ and mass $m_{0}$ (pion mass is assumed for unidentified particles).  
    It is quite natural to describe transverse fragmentation with respect to transverse rapidity in analogy to analysis of longitudinal string fragmentation.}
  where the soft component was localized at
  $y_{t}<2$ ($p_{t} \lesssim 0.5$ GeV) for each particle in the pair, while a Gaussian distribution on $y_{t}$ peaked at $y_{t}\sim2.7$ 
  describes the semi-hard region \cite{pp corr}.  

  Using $y_{t}$ as a cut space, the \emph{angular} correlations for soft and semi-hard components were examined.
  The semi-hard component correlations produce a large peak centered at the $(\eta_{\Delta},\phi_{\Delta})$ origin plus an away-side ridge 
  at $\phi_{\Delta}=\pi$ and are fully consistent with 
  conventional high-$p_{t}$ jet angular correlations, even though the measurements extend to a much lower momentum range and use no leading particle trigger.  
  The observation of jet-like correlations at a lower momentum range underpins the terminology of \emph{semi}-hard scattering and \emph{mini}jet (minimum-bias jet) 
  to distinguish these correlations from those obtained by conventional high-$p_{t}$ triggered particle analysis.
  One surprising result is that minijet correlations in p-p collisions are not symmetric on relative pseudorapidity and azimuth.  
  Instead, the width on $\phi_{\Delta}$ is nearly 50\% larger
  than the width on $\eta_{\Delta}$, in strong disagreement with an angularly symmetric jet cone.  
  The soft component contains a large HBT peak (only observed in like-sign pairs) at the origin
  and a 1D Gaussian centered at $\eta_{\Delta}=0$ that is 
  independent of $\phi_{\Delta}$, consistent with longitudinal string fragmentation.  This two-component model provides a framework for understanding
  correlations in RHIC heavy-ion collisions where the same partonic scattering processes observed in p-p are assumed to be embedded in an unknown medium.
  
  A previous analysis reported angular correlations in Au-Au collisions at 130 GeV \cite{130 CI} (using correlation measure 
  $\bar{N}\frac{\Delta\rho}{\rho_{ref}}$) showing substantial differences between p-p and Au-Au as well as substantial evolution of observed correlation
  structures with centrality.  That analysis accepted a transverse momentum range of $0.15\leq p_{t} \leq 2.0$ GeV/c, containing both soft and semi-hard 
  components.  The soft component, represented by a 1D Gaussian at $\eta_{\Delta}=0$, is diminished relative to p-p in the peripheral centrality bin
  and is not observed in mid-centrality, suggesting that longitudinal string fragmentation is not a relevant mode of particle production in central heavy-ion collisions.
  The semi-hard component contains a minijet cone that in p-p is elongated on $\phi_{\Delta}$, and in Au-Au deforms with increasing
  centrality to become significantly more broad on $\eta_{\Delta}$ and slightly narrower on $\phi_{\Delta}$, reversing the asymmetry found in p-p.  
  The most straightforward interpretation of the elongation in $\eta_{\Delta}$ is to assume that the same semi-hard partonic scattering processes 
  measured in p-p collisions are now embedded in the longitudinally-expanding medium created in heavy-ion collisions.  
  Multiple hadronic scattering and other final-state interactions would most likely broaden the minijet peak 
  along both $\eta_{\Delta}$ and $\phi_{\Delta}$, and are unlikely to be of sufficient magnitude to explain this effect.  However, partonic interactions with
  a longitudinally-expanding medium could be the source of the observed asymmetry reversal.

\begin{figure}[p]
    \includegraphics[width=1.0\textwidth]{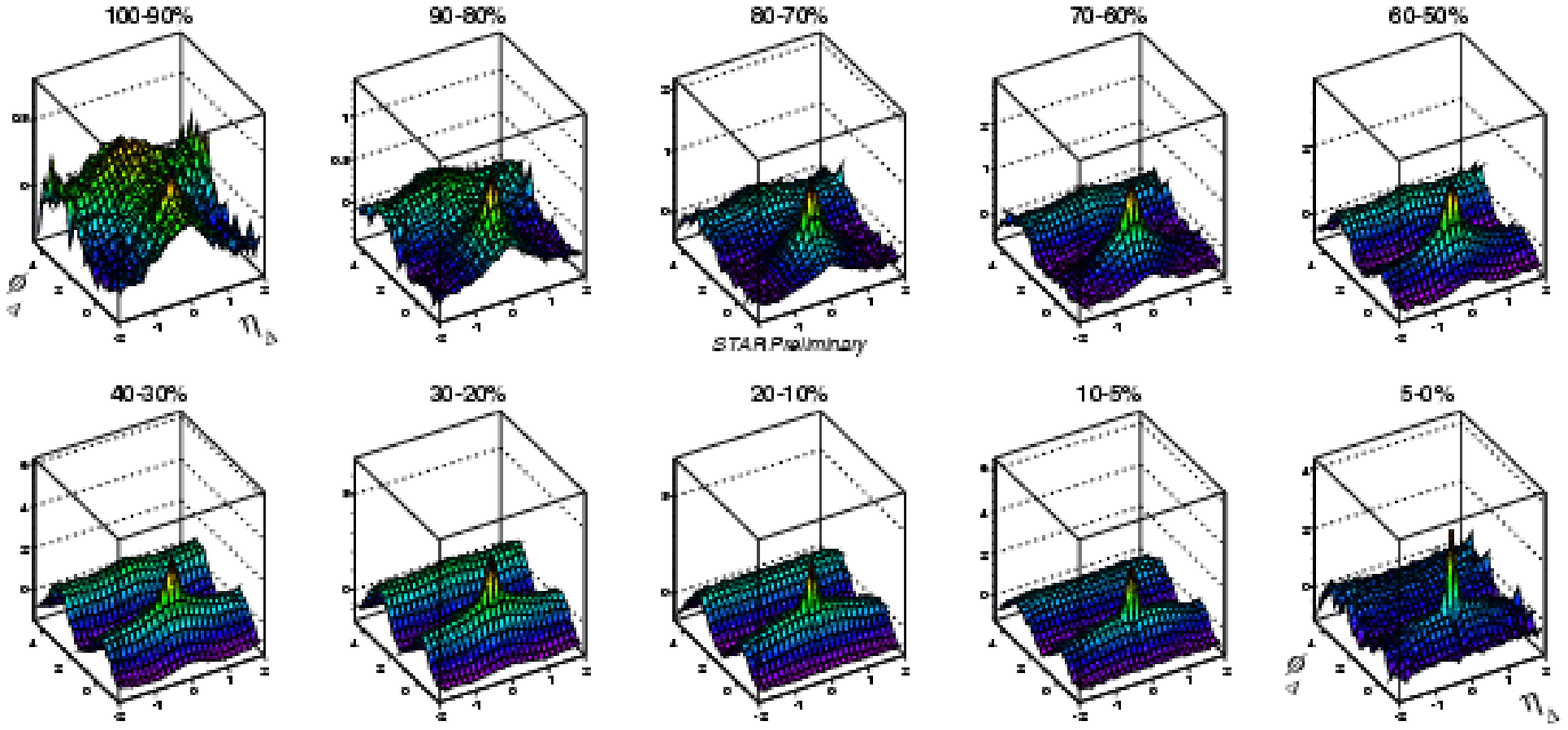}
    \caption{Au-Au 62 GeV CI angular correlations.  Note changes in scale.  See text for discussion.}
    \label{62 CI}
  \end{figure}

\begin{figure}[p]
    \includegraphics[width=1.0\textwidth]{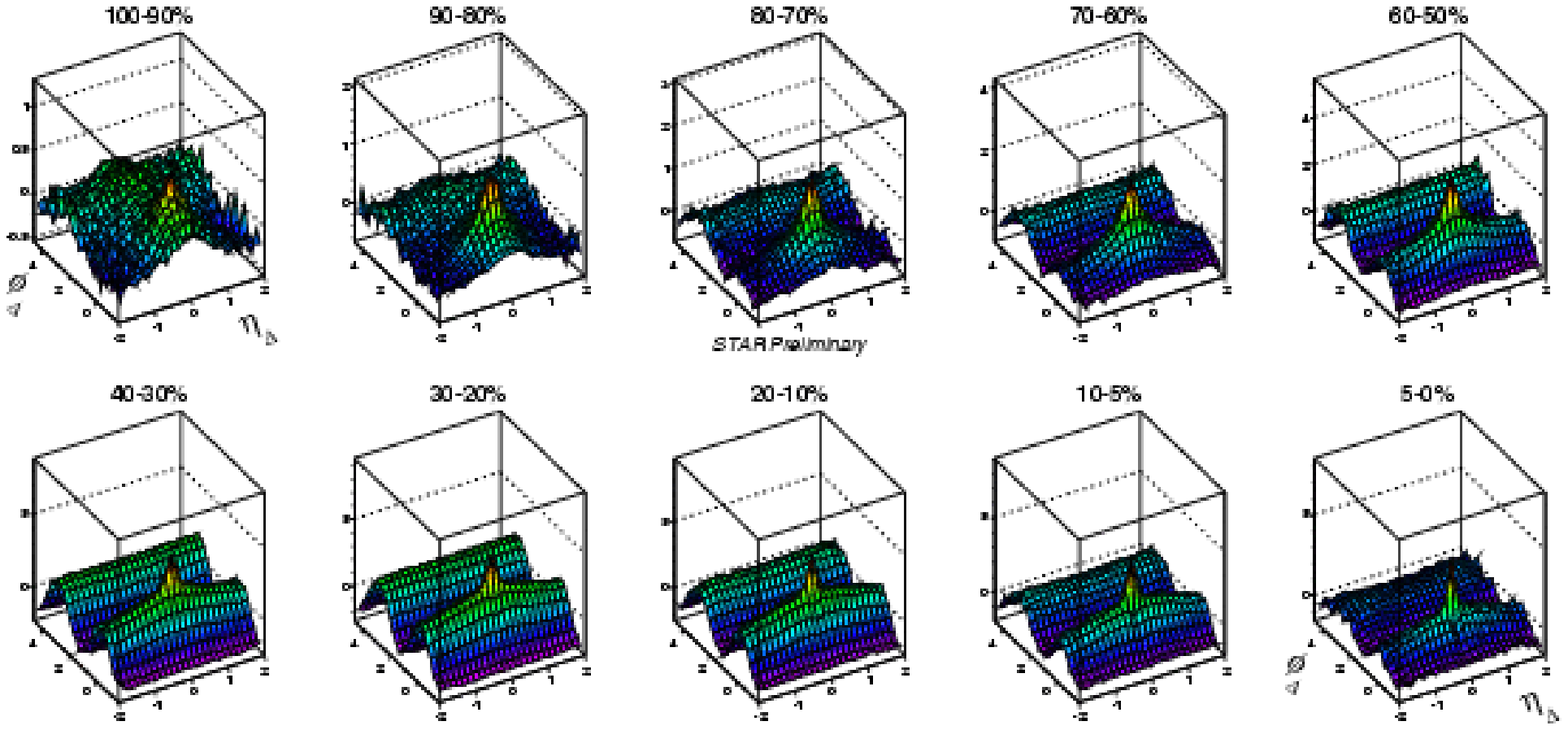}
    \caption{Au-Au 200 GeV CI correlations.}
    \label{200 CI}
  \end{figure}

  Unfortunately, the limited volume of the 130 GeV Au-Au data
  only allowed four centrality bins, the most peripheral being 40-70\% of the total cross section, and could not reveal the full
  centrality dependence of correlation structures and the transition in peripheral collisions from p-p to Au-Au.
  The present study analyzes nearly an order of magnitude more events at both 62 and 200 GeV which are divided into eleven centrality bins spanning the entire range 
  from 0-100\% of total
  cross section, revealing for the first time the detailed centrality and energy dependences of these correlation structures, as shown in Figures
  \ref{62 CI} and \ref{200 CI}.  The data span the transverse momentum range $0.15\leq p_{t} \leq 6.0$ GeV/c and, as in the 130 GeV analysis, 
  contain no trigger particle of correlation model assumption.

  These correlation structures may be summarized by three general trends.  First, the most peripheral (90-100\%) centrality bins are in excellent agreement with p-p data, and 
  the correlation structures evolve smoothly with increasing centrality for each energy.  Second, the soft component (1D Gaussian at $\eta_{\Delta}=0$) 
  diminishes with increasing centrality and does so much more rapidly with increasing energy.  The soft component is visible at 62 GeV until $\sim$20\% 
  centrality whereas at 200 GeV it is completely absent by 70\%.  Finally, the hard component minijet peak is dramatically broadened in $\eta_{\Delta}$ with
  increasing centrality and this deformation occurs much more rapidly at higher energy.  

  The HIJING Monte Carlo simulation model includes particle production from soft fragmenting longitudinal strings and hard perturbative QCD jet production and 
  fragmentation with jet quenching.
  The results from HIJING simulations qualitatively agree with the peripheral data, though HIJING substantially over-predicts the amount of soft component
  relative to hard and fails to predict the energy and centrality dependence.
  HIJING also fails to predict the asymmetry of $\eta_{\Delta}$ and $\phi_{\Delta}$, which is unaffected by the jet quenching option, and actually predicts 
  very little change with centrality.  These results suggest that the observed broadening along $\eta_{\Delta}$ may be due to microscopic interactions with the medium
  not modeled by HIJING. Future studies including partonic interactions within the AMPT model will examine this effect in more detail.



  \section{Charge-dependent correlations}

  The above analysis can be extended to differentiate between the relative electric charge sign of particles.  Defining like-sign pairs (LS: $++$,$--$) and 
  unlike-sign pairs (US: $+-$,$-+$) the above charge-independent (CI) correlation is formed as CI = LS + US.  A charge-dependent (CD) correlation is defined
  as CD = LS - US, where the sign convention is chosen to be compatible with isospin conventions\footnote{The isovector term in the two-particle density is the
    coefficient of $\tau_{1} \cdot \tau_{2}$ where $\tau$ is the z-component of isospin.  
    For a pair of LS charged pions $\tau_{1}\cdot\tau_{2} = +1$ while US $\tau_{1}\cdot\tau_{2} = -1$, thus CI = LS + US isolates the isoscalar component, 
    while CD = LS - US is isovector.}.
  Charge-dependent angular correlations in p-p collisions are dominated by a negative 1D Gaussian on $\eta_{\Delta}$ with width $\sigma_{\eta\Delta} \simeq 1$
  \cite{pp CD, ACCDHW}.  This feature is conventionally attributed to charge-ordering during longitudinal string fragmentation \cite{string}
  and shows the dependence of the soft component on relative charge sign.  
  Since HBT correlations are only present in LS pairs, they form the only other significant correlation feature of a positive Gaussian peak centered at the origin.

  In central Au-Au collisions at 130 GeV the 1D Gaussian is not detected, instead a negative \emph{2D} exponential dominates the correlation structure,
  suggesting charge-ordering is no longer occurring along a 1D string as in p-p but along a 2D surface \cite{130 CD}.  The change from Gaussian to exponential
  distributions may be caused by pair attenuation which increases with opening angle due to interactions with a the medium.   
  Lower centralities show intermediate stages between p-p and central Au-Au, but again this analysis only shows coarse centrality dependence.  

\begin{figure}[p]
    \includegraphics[width=1.0\textwidth]{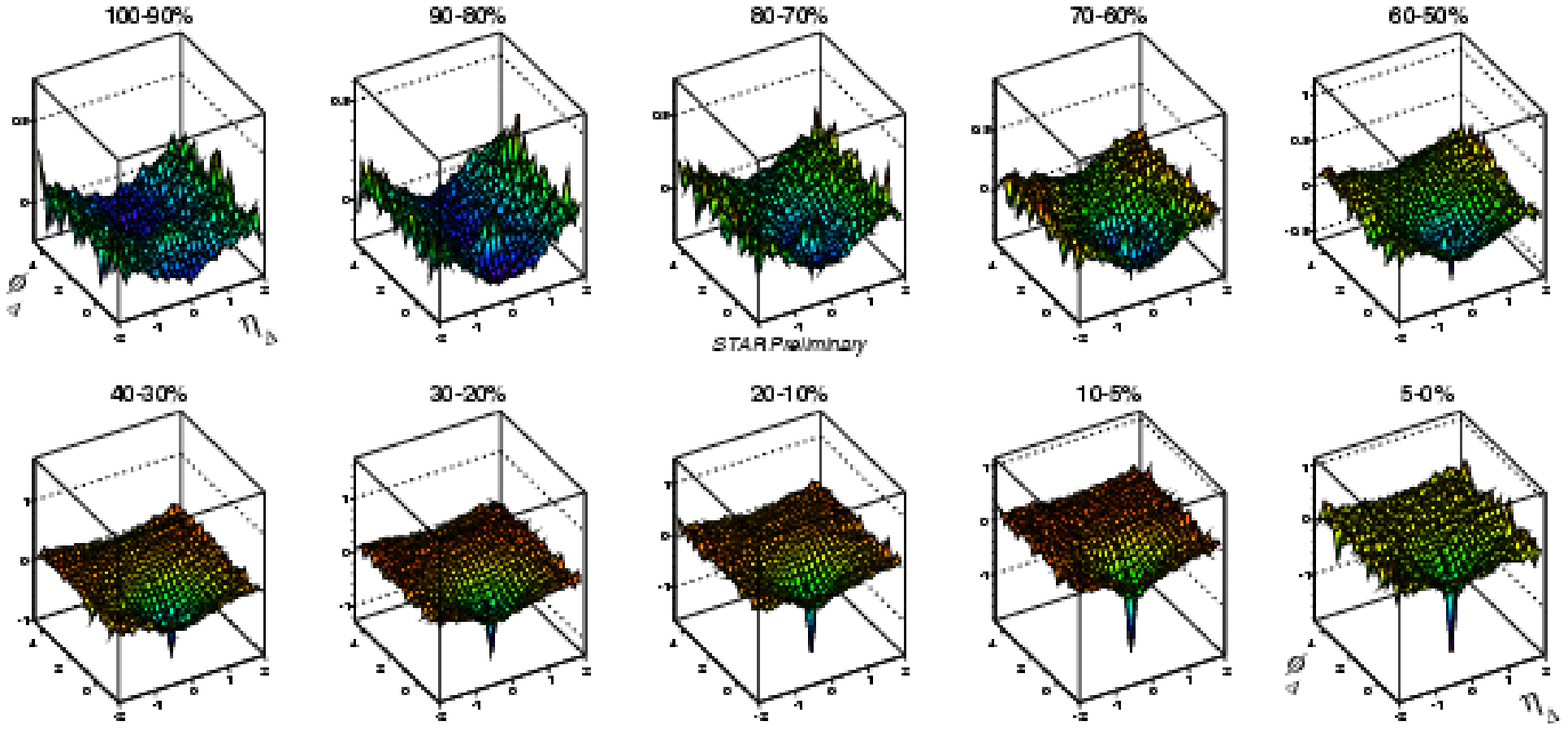}
    \caption{Au-Au 62 GeV CD angular correlations.  Note changes in scale.  See text for discussion.}
    \label{62 CD}
  \end{figure}

\begin{figure}[p]
    \includegraphics[width=1.0\textwidth]{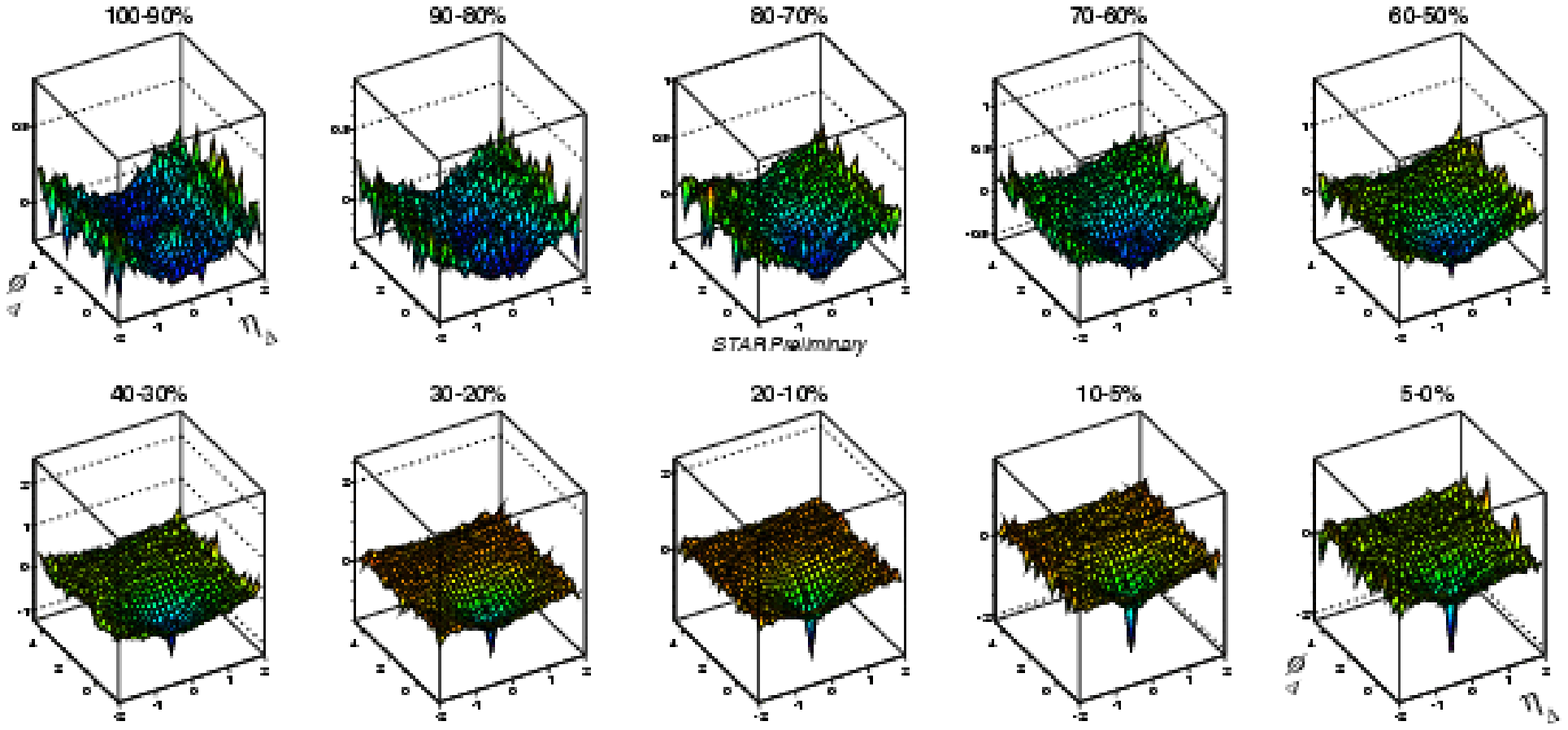}
    \caption{Au-Au 200 GeV CD correlations.}
    \label{200 CD}
  \end{figure}

  The present analysis shows angular CD correlations in Au-Au collisions at 62 and 200 GeV in Figures \ref{62 CD} and \ref{200 CD}.  As in the previous section, excellent
  agreement between the most peripheral (90-100\%) centrality bin and p-p is demonstrated and the correlation structure evolves smoothly with increasing centrality.
  These results confirm the independent observation at 130 GeV that the 1D Gaussian on $\eta_{\Delta}$ in p-p, attributed to longitudinal charge-ordering, is
  replaced by a 2D exponential approximately symmetric on $\eta_{\Delta}$ and $\phi_{\Delta}$.  
  The evolution occurs more rapidly at higher energy with a more dramatic decrease in widths of the exponential.

  The primary source of CD correlations appears to be charge-ordering due to local charge conservation in fragmentation processes into quark-antiquark pairs.  
  It is more likely for neighboring particles to have to opposite charge, producing a stronger correlation in US than LS, thus the CD correlation is negative.
  The symmetry axis of charge-ordering is telling of the fragmentation process.  In elementary collisions charge-ordering can be seen on the longitudinal
  ($z$ or $\eta$) direction from string fragmentation phenomenologically described by the Lund model \cite{string}.  It is also possible to observe 
  transverse charge-ordering associated with jet fragmentation processes on transverse rapidity, which are strikingly similar to CD correlations of soft 
  longitudinal fragmentation on longitudinal rapidity.

  In Au-Au collisions at RHIC, we observe a significant evolution of the charge-ordering signal.  Peripheral CD correlations, which are in excellent 
  agreement with p-p data, show a 1D charge-ordering signal on $\eta_{\Delta}$ as expected from fragmentation of 1D longitudinal strings.  
  HIJING predicts no change in CD correlations with
  energy or centrality, but the data show otherwise.  We observe that 1D charge-ordering on $\eta_{\Delta}$ evolves to become 2D charge-ordering on $\eta_{\Delta}$ 
  and $\phi_{\Delta}$, measuring \emph{for the first time} charge-ordering in the azimuthal direction.  These results suggest that in central Au-Au collisions
  particle production does not occur along several independent 1D longitudinal strings, but along a surface in at least two dimensions (the transverse direction
  is not considered here).  Additionally, we also find a change in shape with increased centrality of the charge-ordering signal from Gaussian to exponential.  
  This change is expected in a scenario where rescattering increases with the opening angle of the pair.  While hadronic rescattering may contribute to this effect, 
  it is extremely unlikely that rescattering alone could cause a 1D signal to appear to be both 2D and symmetric without invoking new hadronization mechanisms.

  Several net-charge fluctuation measures have been proposed and studied to search for QGP formation \cite{net-charge}.  Those analyses have shown little or no centrality
  dependence in sharp contrast to the predictions of dramatic suppression of net-charge fluctuations.  We must again emphasize that net-charge fluctuation
  measures typically integrate over the correlations shown here and are therefore insensitive to changes in differential structure.  
  It is significant to note that while we observe a dramatic change in the amplitude, width, and shape of charge-ordering correlations with centrality, 
  the integrals of these correlations exhibit only modest changes with centrality (at 130 GeV the integrals increase by about 20\% from peripheral to central \cite{130 CD}).


  \section{Summary}

  The primary goal of the heavy-ion physics program at RHIC is to determine if the nuclear matter created in RHIC collisions is of sufficient energy
  density such that the relevant degrees of freedom are partonic rather than hadronic and, if this partonic medium is created, to measure its properties.
  To that end we have presented a general analysis technique based on minimum-bias two-particle correlations which are shown to be sensitive to the low $Q^2$ kinematic
  range where scattered partons strongly interact with the bulk medium.  In addition to providing access to 
  new physics, this analysis represents a generalization of both triggered jet correlations and event-by-event
  fluctuation analyses, enabling results from these methods to be evaluated in a much larger context and from the point-of-view of a more complete physical picture.

  Surveying the correlations in Au-Au collisions at 62 and 200 GeV, along with previous analyses at 130 GeV and in proton-proton collisions, 
  provides an overview of the centrality and energy dependence of correlation structures from heavy-ion to nucleon-nucleon collisions.  
  Charge-independent correlations reveal the attenuation of soft longitudinal string fragments and dramatic broadening of the minijet correlation peak,
  both phenomena are consistent with interactions with a partonic medium developing with increasing centrality and collision energy.
  Charge-dependent correlations suggest a fundamental change in hadronization geometry from one dimensional longitudinal strings to a multi-dimensional
  surface, also consistent with the hadronization of a partonic medium.


  I thank the organizers of the 2nd International Workshop on Correlations and Fluctuations in Relativistic Nuclear Collisions. 
  This research in supported in part by the U.S. Department of Energy.

\end{document}